\pdfoutput=1
\documentclass[aps, reprint, twocolumn,longbibliography,nofootinbib,floatfix]{revtex4-1}
\usepackage[latin1]{inputenc}
\usepackage{graphicx}
\usepackage{amsmath,amssymb,amstext,graphicx,bm,booktabs,hyperref}
\usepackage{fancyhdr}

\newcommand*{\affaddr}[1]{#1}
\newcommand*{\affmark}[1][*]{\textsuperscript{#1}}
\fancypagestyle{alim}{\fancyhf{}\fancyfoot[R]{$^\dagger$Both authors contributed equally and are displayed in alphabetical order.}}

\begin{document}

\title{Efficiently measuring a quantum device using machine learning}

\author{D.T. Lennon\affmark[$\dagger$1], H. Moon\affmark[$\dagger$1], L.C. Camenzind\affmark[2], Liuqi Yu\affmark[2], D.M. Zumb\"uhl\affmark[2], G.A.D. Briggs\affmark[1], M.A. Osborne\affmark[3], E.A. Laird\affmark[4], and N. Ares\affmark[1]\\
\affaddr{\affmark[1]Department of Materials, University of Oxford, Parks Road, Oxford OX1 3PH, United Kingdom}\\
\affaddr{\affmark[2]Department of Physics, University of Basel, 4056 Basel, Switzerland}\\
\affaddr{\affmark[3]Department of Engineering, University of Oxford, Walton Well Road, Oxford OX2 6ED, United Kingdom}\\
\affaddr{\affmark[4]Department of Physics, Lancaster University, Lancaster, LA1 4YB, United Kingdom}
}

\begin{abstract}
Scalable quantum technologies will present challenges for characterizing and tuning quantum devices. This is a time-consuming activity, and as the size of quantum systems increases, this task will become intractable without the aid of automation. We present measurements on a quantum dot device performed by a machine learning algorithm. The algorithm selects the most informative measurements to perform next using information theory and a probabilistic deep-generative model, the latter capable of generating multiple full-resolution reconstructions from scattered partial measurements.
We demonstrate, for two different measurement configurations, that the algorithm outperforms standard grid scan techniques, reducing the number of measurements required by up to 4 times and the measurement time by 3.7 times. 
Our contribution goes beyond the use of machine learning for data search and analysis, and instead presents the use of algorithms to automate measurement. This work lays the foundation for automated control of large quantum circuits. 
\end{abstract}
\date{\today{}}
\maketitle
\thispagestyle{alim}
Semiconductor quantum devices hold great promise for scalable quantum computation. In particular, individual electron spins in quantum dot devices have shown long spin coherence times with respect to typical gate operation times, high fidelities, all-electrical control, and good prospects for scalability and integration with standard semiconductor technologies~\cite{Vandersypen2017}.

A crucial challenge of scaling spin qubits in quantum dots is that electrostatic confinement potentials have large variability among devices and even in time, predominately due to charge traps. The characterisation of such devices, which implies measurements of current or conductance at different applied biases and gate voltages, can be very time consuming. It would normally be carried out following simple scripts such as a grid scan, which is inefficient and slow. Optimising this process involves selecting those biases and gate voltages for which measurements are more informative. An optimised method for device measurement is key to automate device tuning. Current efforts towards automating quantum dot tuning are based on grid scans and require several hours to execute, lack generality across devices, and/or require manual input~\cite{Baart2016,Stehlik2015,Kalantre2017}. 

\begin{figure}[!t]
\centering
\includegraphics[width=0.5\textwidth]{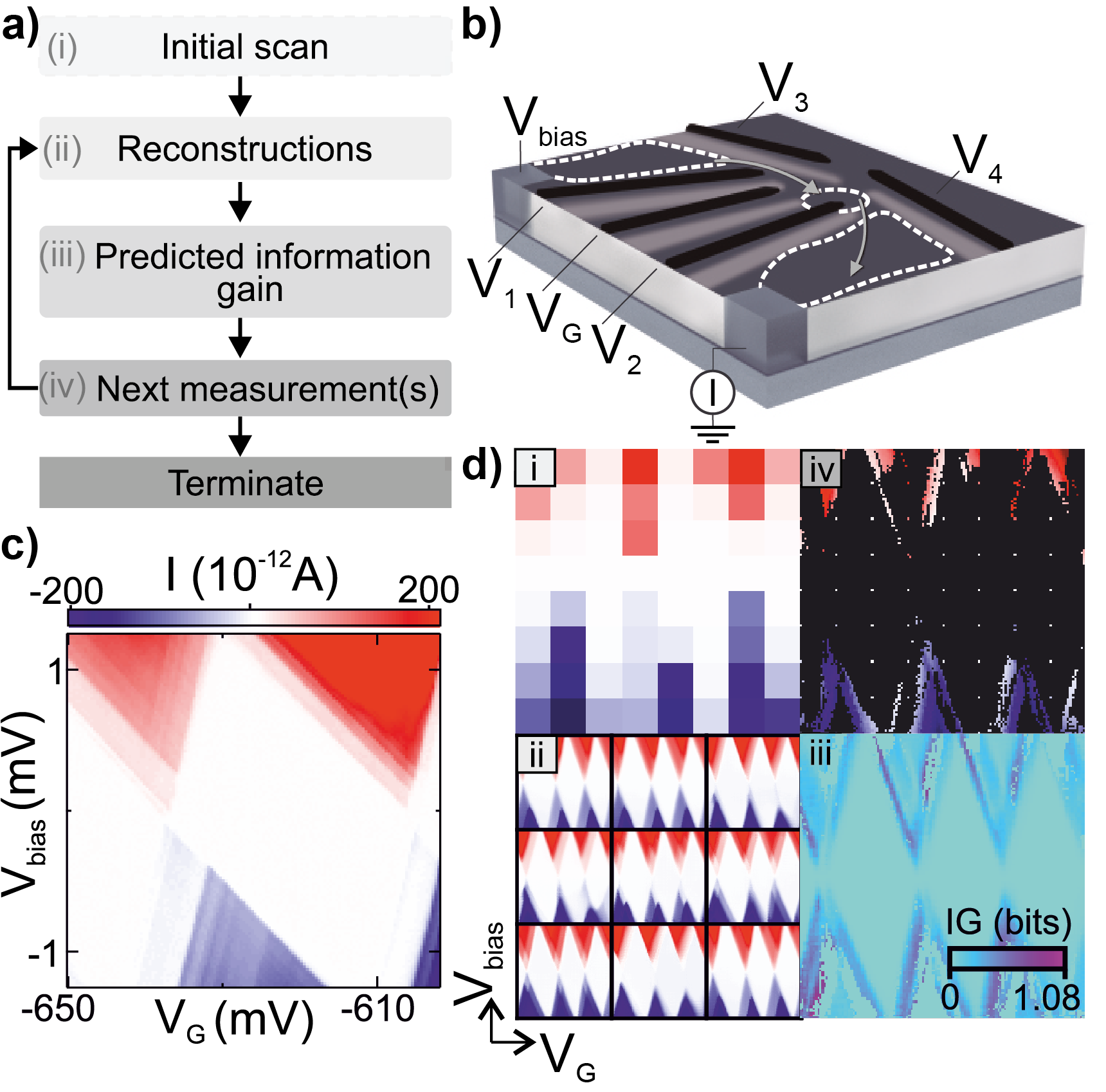}
\caption{\label{Fig1}
Overview of the algorithm and the quantum dot device. (a) Schematic of the algorithm's operation. Low-resolution measurements (i) are used to produce reconstructions (ii), which are used to infer the predicted information gain acquisition map (iii). Based on this map, the algorithm chooses the location of the next measurement (iv). The process is repeated until a stopping criterion is met. (b) Schematic of the device. A bias voltage $V_{\text{bias}}$ is applied between ohmic contacts to the two-dimensional electron gas. We apply gate voltages labelled $V_1$ to $V_4$ and $V_\text{G}$. (c) A measured current map as a function of $V_{\text{bias}}$ and $V_\text{G}$. The Coulomb diamonds are the white regions where electron transport is suppressed, and most of the information necessary to characterize a device is contained just outside these regions. (d) Sequential decision algorithm in (a) illustrated with an example of a specific current map. In panel (iv), unmeasured pixels are plotted black; however, initial measurements (i) are represented so as to fill the entire panel (that is, the sparse grid of measurements is represented as a low-resolution image).}
\end{figure}

In this paper, we present an algorithm that performs efficient real-time data acquisition for a quantum dot device. It starts from a low-resolution uniform grid of measurements, creates a set of full-resolution reconstructions, calculates the predicted information gain (or acquisition map), and selects the measurements which will give the maximum information gain (Fig.~\ref{Fig1}a). This process is iterated until the information gain from new measurements is marginal.

Such an information-theoretic criterion for selecting measurements is based on an uncertainty measure of random variables~\cite{Houlsby2011,Ankenman2010,Sacks1989}, and hence a probabilistic model is required on the unobserved variables. Rather than using a Gaussian process~\cite{Rasmussen2005}, we use a conditional variational auto-encoder (CVAE)~\cite{Sohn2015}, which is capable of generating high-resolution reconstructions given partial information and is fast enough for real-time decisions. In spite of their suitability, these models have not previously been applied to efficient data acquisition. Deep generative models such as the variational auto-encoder (VAE)~\cite{Kingma2014} and generative adversarial networks (GAN)~\cite{Goodfellow2014} have shown great success in generating complex non-stationary patterns of data and multi-modal distributions~\cite{Mescheder2017,Srivastava2017}. Deep generative models are used for: speech synthesis~\cite{Oord2016_Wavenet}; generating images of digits and human faces~\cite{Makhzani2015}; transferring image style~\cite{Zhu2017,Taigman2017}; and inpainting missing regions of images~\cite{IizukaSIGGRAPH2017}. Recently, VAE models have been used in scientific research to optimise molecular structures~\cite{Sanchez2018,Gomez2018,Kusner2017,Dai2018}. An advantage of deep generative models over simple interpolation techniques, such as nearest-neighbour and bilinear interpolation, is that deep generative models can learn likely patterns from training data and utilise such patterns to make reconstructions.

Our device is a laterally defined quantum dot fabricated by patterning Ti/Au gates over a GaAs/AlGaAs heterostructure containing a two-dimensional electron gas (Fig.~\ref{Fig1}b). In this device, electrons are subject to the confinement potential created electrostatically by gate voltages. Gate voltages $V_1$ to $V_4$ tune the tunneling rates while $V_\text{G}$ mainly shifts the electrical potential of the dot level. The current through the device is determined both by these gate voltages and by the bias voltage $V_{\text{bias}}$. Our measurements were performed at $30$~mK. In Fig.~\ref{Fig1}c we show a current map as a function of $V_\text{G}$ and $V_{\text{bias}}$ for fixed values of $V_1$ to $V_4$. Diamond shaped regions or `Coulomb diamonds', correspond to Coulomb blockade, where electron tunnelling is suppressed~\cite{Hanson2007}. Most current maps have large areas in which the current is almost constant, and consequently measurements in these regions slow down informative data acquisition dramatically. The device current gradient is the derivative of the current with respect to bias and gate voltages, and therefore regions of high current gradient are typically very informative for device characterization. Our algorithm gives measurement priority to the informative regions of the current map, which leads to measurements that concentrate in regions of high current gradient. An overview of an algorithm-assisted measurement of a current map is shown in Fig.~\ref{Fig1}d.

\section*{Reconstruction model and training}
The role of the reconstruction model is to characterise likely patterns in a training dataset, derived from a mixture of measured and simulated current maps. We can utilise these likely patterns to predict the unmeasured signals from given partial measurements.   

Deep generative models represent this pattern characterisation in a low-dimensional real-valued latent vector $z$, which can be decoded to produce a full-resolution reconstruction. The latent space representation and the decoding are learned during training. The CVAE that we use consists of two convolutional neural networks, an encoder and a decoder. The encoder is trained to map full-resolution training examples of current maps $Y$  to the latent space representation $z$. 
 
The encoder also enforces that the distribution $p(z)$ of training examples in latent space is Gaussian.
The decoder is trained to reconstruct $Y$, from the representation $z$ combined with an $8\times8$ subsample of $Y$. As a result, $z$ attempts to represent all the information that is missing from the subsampled data. The chosen loss function, which the CVAE attempts to minimise, is a measure of the difference between the training data and the corresponding reconstruction. To avoid blurry reconstructions, we define a contextual loss function that incorporates both pixel-by-pixel and higher-order differences like edges, corners, ans shapes. Detailed description of these networks and their training can be found in the Supplementary Information.

The model is trained using both simulated and measured current maps. We choose to work with current maps of resolution $128\times128$. The simulation is based on a constant-interaction model (see Methods). To measure the current maps for training, we set the bias and gate voltages ranges randomly from a uniform distribution. The training dataset consists of 25,000 simulations and 25,000 real examples generated by randomly cropping 750 measured current maps. The current maps were subjected to random offsets, rescaling, and added noise to increase the variability of the training set. 

\section*{Generating reconstructions from partial data}
The trained decoder network is now used in the algorithm of Fig.~\ref{Fig1}a to reconstruct full-resolution current maps from partial data. At each stage, the known partial current map is denoted $Y_n$, where $n \leq 128^2=16,384$ is the number of measured pixels. To generate a reconstruction, the decoder takes as input the initial $8\times 8$ grid scan $Y_{64}$, together with a latent vector $z$ sampled from the posterior distribution $p(z|Y_n)$. The latent space of $z$ and the prior probability $p(z)$ are constructed by the CVAE during training, but the posterior distribution takes account of all $n$ measurements (for details, see Methods). The posterior samples are drawn from $p(z|Y_n)$ by the Metropolis-Hastings (MH) method, of which one iteration moves previous samples towards $p(z|Y_n)$. More iterations make better samples of $p(z|Y_n)$. The samples of $z$ are then converted to $\hat{Y}_m$, where $m=1,\ldots,100$ is the reconstruction index. The continuous posterior $p(z|Y_n)$ is then approximated by a discrete posterior of samples $P_n(m)$, which denotes how probable $\hat{Y}_m$ is. We refer to $P_n(m)$ as the posterior distribution of reconstructions.

\section*{Sequential measurement decision}
\begin{figure}
\centering
\includegraphics[width=0.5\textwidth]{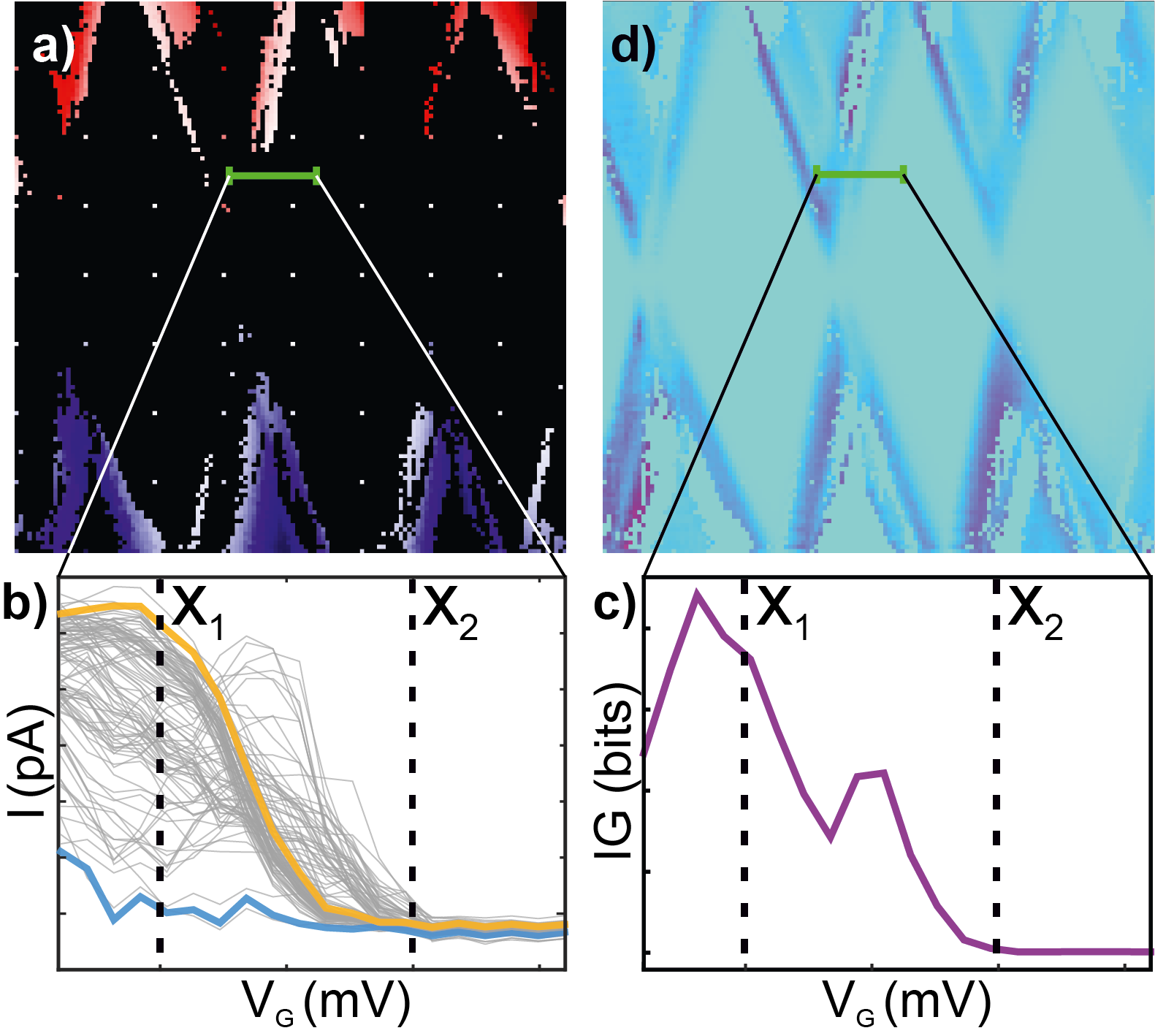}
\caption{\label{Fig2}
Computation of the acquisition map. (a) Partial current map. To illustrate the first step in the computation of the acquisition map, we consider a trace (green) through an unmeasured region of the map. (b) For the unmeasured trace in (a), reconstructions provide $M$ different predictions. Blue and yellow traces highlight two of these predictions. The objective is to determine the most informative measurement location. At $x_2$, all predictions are similar, so measuring here will have little impact on the posterior distribution of reconstructions. At $x_1$, predictions are dissimilar and this is therefore a more informative measurement location, with a larger effect on the posterior distribution of reconstructions. (c) Information gain computed for the unmeasured trace in (a). (d) Acquisition map of information gain computed from the partial measurements in (a), and plotted over the entire image range.
}
\end{figure}

With each iteration of the decision algorithm, an acquisition map is computed from the accumulated partial measurements and the resulting reconstructions. The purpose of this acquisition map is to indicate how informative potential measurement locations are for the posterior distribution of reconstructions (Fig \ref{Fig2}). The (n+1)th measurement, whose result is $y_{n+1}$ is one pixel taken from the true current map, changes our posterior distribution from $P_n(m)$ to $P_{n+1}(m)$, rendering different reconstructions more or less probable.

The acquisition map is the expected information gain $IG(x)$ at each potential measurement location $x$.
Our algorithm calculates it by a weighted sum over reconstructions:
\begin{equation}
IG(x)\equiv\sum_{m} P_n(m) \times IG_m \bigl( x \bigr) \quad ,
\end{equation}
where $IG_m(x)$ is the Kullback-Leibler divergence between the distributions $P_n$ and $P_{n+1}$, calculated under the assumption that $y_{n+1}$ is taken at location $x$ from reconstruction $\hat{Y}_m$.
The most informative point is $x_{n+1}^* \equiv \mathrm{argmax}_x IG(x)$. This criterion is equivalent both to minimising the expected information entropy of the posterior distribution and to Bayesian active learning by disagreement (BALD)~\cite{Houlsby2011} (see Methods). 

We devised a choice of two methods to make decisions based on the acquisition map; a pixel-wise method, and a batch method. The pixel-wise method selects the single best location in the acquisition map. This method is not optimal in terms of measurement time, as it might collect data from locations that require a large gate voltage ramp. The ramp rate is limited by the measurement electronics and the device settling time. The batch method selects multiple locations from the acquisition map, and then acquires selected measurements taking into account the distance between locations, thus reducing the measurement time compared with the pixel-wise method. 

\section*{Results}
\begin{figure}
\includegraphics[width=0.5\textwidth]{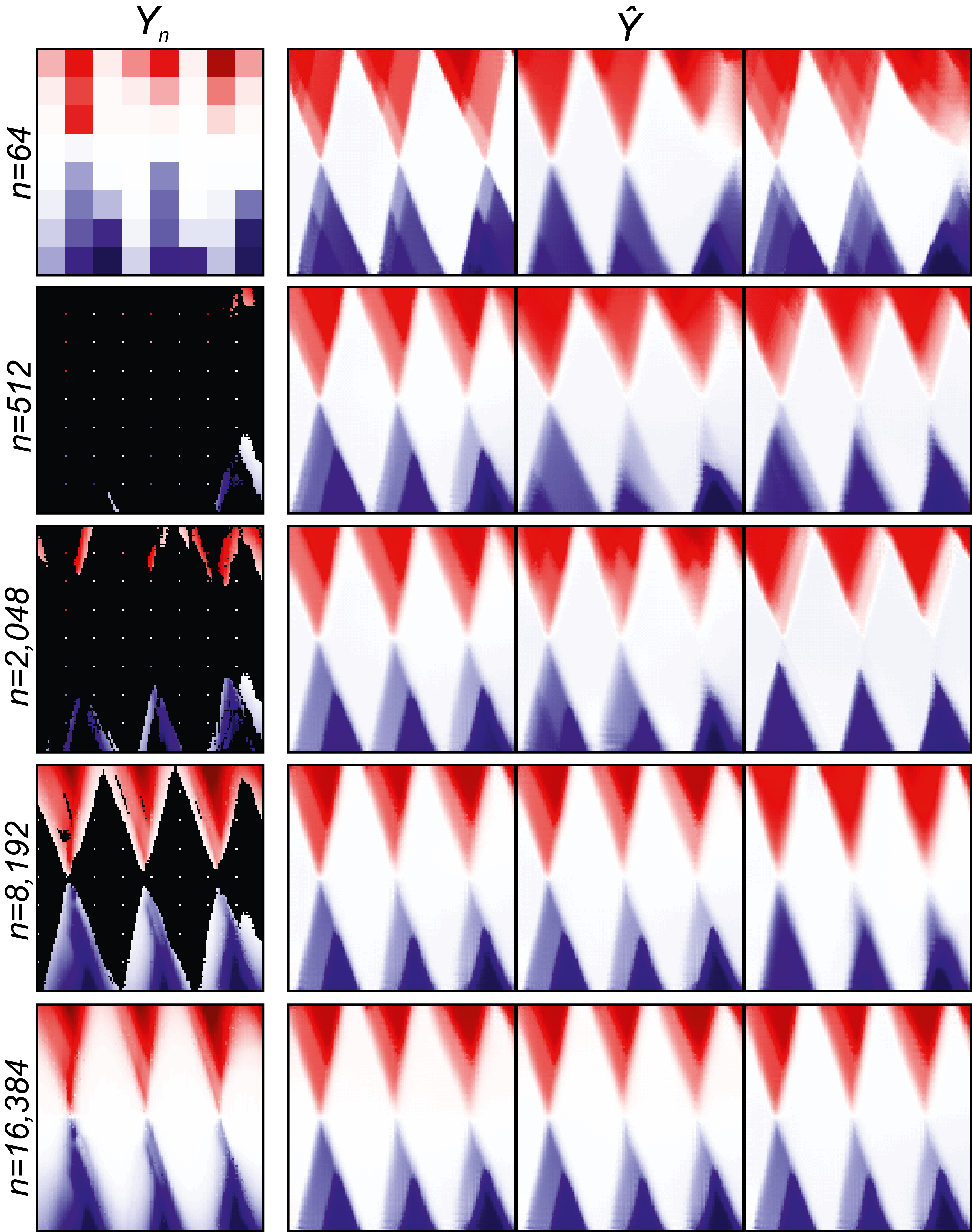}
\caption{\label{Fig3}
Posterior update of reconstructions. In each row, the first column shows the algorithm-assisted measurements, using the batch method, for a given $n$. The remaining three columns contain example reconstructions given the corresponding $n$ measurements. As $n$ increases, the diversity of the reconstructions is reduced and their accuracy increased. There is still uncertainty remaining even in the last row -- the posterior distribution still contains variance.
}
\end{figure}
To test the algorithm, it was used to acquire a series of current maps in real time. First, the device was thermally cycled, to randomise the charge traps and therefore present the algorithm with a configuration not represented in its training data. Next, gate voltages $V_1$ -$V_4$ were set to a new combination of values, and the algorithm was tasked to measure the corresponding current map using both the batch and the pixel-wise methods. This step was repeated for ten different voltage combinations. Fig \ref{Fig3} presents data acquired during a typical iteration, together with selected reconstructions at each stage. As expected, reconstructions become less diverse as more measurements are acquired. The reconstructions do not necessarily replicate the measured current map for large $n$. This is because reconstructions have a limited variability given by the training data.
Decisions are made based on the learned patterns from the training data, which implies that the training data should contain at least general patterns which are to be characterised. Consequently, the training dataset does not need to include all possible features in a current map.

\begin{figure*}
\includegraphics[width=\textwidth]{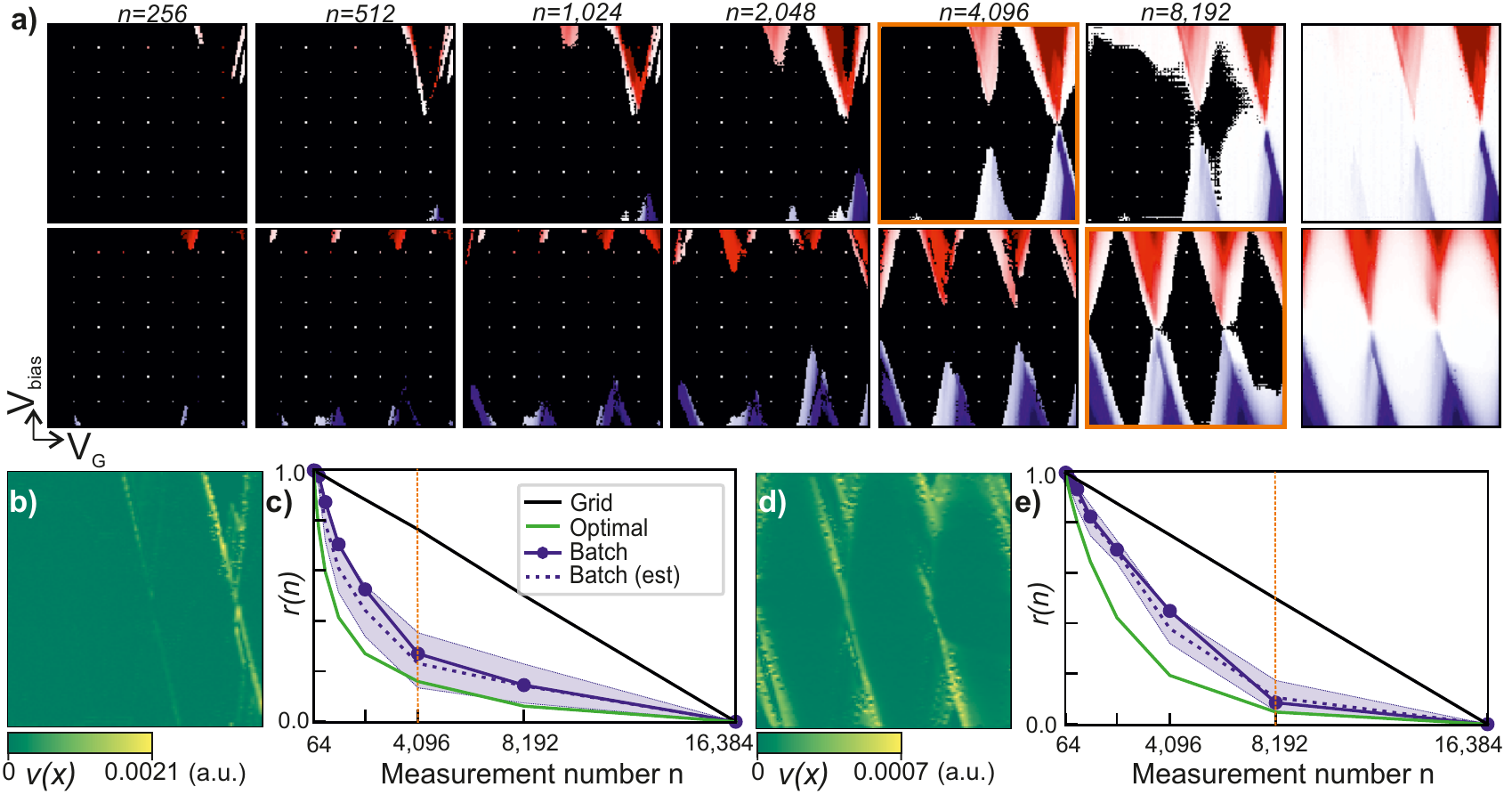}
\caption{\label{Fig4}
Algorithm-assisted measurements of Coulomb diamonds. (a) Sequential batch measurement in two different experiments. Each row displays algorithm assisted measurements of the current map as a function of $V_{\text{bias}}$ and $V_\text{G}$ for different values of $n$. The last plot in each row is the full-resolution current map. (b,d) Current gradient map for each example in (a). (c,e) Measure of the algorithm's performance $r(n)$, average real-time estimate of $r(n)$ across reconstructions with 90\% credible interval, and optimal $r(n)$ for both examples in (a). The black line is the value of $r(n)$ corresponding to the alternating grid scan method. The dashed orange line indicates the value of $n$ determined by the stopping criterion. 
}
\end{figure*}

In Fig.~\ref{Fig4}a two representative measurement sequences using the batch method are shown. The algorithm avoids measurements in regions of low current gradient. These regions coincide with the interiors of the Coulomb diamonds for the cases considered. This strategy is an emergent property of the algorithm and is wise; little information about the device characteristics can be found in low-current gradient regions of the current map. This preference derives from the comparison between reconstructions, which exhibit the greatest disagreement outside Coulomb diamonds. The performance for other current maps, and for the pixel-wise method, are shown in the Supplementary Information.

We compared the performance of the algorithm with an alternating grid scan method. This type of grid scan starts with 8$\times$8 measurements and alternately increases the vertical and the horizontal grid size by 2 (i.e. 16$\times$8, 16$\times$16, 32$\times$16, etc.), without performing the same measurement twice. Over the ten different current maps, the average time for full-resolution data acquisition with the alternating grid scan method is 554 seconds. This time is limited by our bias and gate voltage ramp rate and chosen settling time. 
The batch method can be implemented with any batch size however for direct comparison with the alternating grid scan we selected increasing batches of 32$\times 2^{b}$, where $b$ is the batch number starting from~1.

Two types of computation are required to make a measurement decision: sampling reconstructions using MH and constructing the acquisition map. One MH sampling iteration takes 63~ms. For experiments, multiple sampling iterations are performed when $n$ reaches one of batch decision points while measurement is suspended. Since, sampling can be performed simultaneously with periods of measurement acquisition, and thus does not add to the measurement time, our reported measurement times in this paper exclude the time for sampling. To compute a single acquisition map takes approximately 50~ms using a NVIDIA GTX 1080 Ti graphics card and Tensorflow~\cite{Tensorflow} implementation. The acquisition map must be computed for every batch or every pixel measurement, except the initial $8\times8$ grid scan and the final acquisition step (which has no choice which pixel(s) to measure). To acquire a full resolution current map thus requires 7 computations (350~ms) for the batch method, and 16,319 computations (816~s) for the pixel-wise method. For the batch method, the computation time is negligible compared to the measurement time, but for the pixel-wise method it is a limiting factor in the measurement rate.

We have developed a measure of the algorithm's performance that is based on the observation that measurements in regions of low current gradient are less informative than regions of high current gradient. Let $v(x)$ denote the numerically approximated euclidean norm of the gradient $\lVert \nabla Y(x)\rVert_2$, which is equivalent to the norm of the current gradient at $x$. Then the error is defined as $r(n) = 1.0 - \frac{V(n)}{V(N)}$, where $N$ is the total number of pixels (in our case 16,384), and $V(n)=\sum_{i=1}^n v(x_i)$ in which $x_i$ is the location of the $i$th measurement. Hence $r(n)$ is the ratio of total current gradient at unmeasured locations to total current gradient in the entire map.
This error can only be calculated after all measurements have been performed.  
However, we can utilise the $m$th reconstruction to generate an estimate $\tilde{r}_m(n)$ in real-time by replacing $\lVert \nabla Y(x) \rVert_2$ with $\lVert \nabla \hat{Y}_m(x)\rVert_2$. The estimates from multiple reconstructions yield a credibility interval for $r(n)$.
The value of $r(n)$ for an optimal algorithm is $\bar{r}(n)=1.0-\frac{V^*(n)}{V(N)}$, where $V^*(n)$ is the sum of largest $n$ values of $v(x)$. This is the performance that would be obtained if each measurement location were chosen knowing the full-resolution current map, and thus which is the next measurement location that corresponds to the highest unmeasured current gradient. No decision method can exceed this bound.
For the real time estimates of $r(n)$, we have increased the number of reconstructions to 3,000 by adding different noise patterns that are present in typical measured current maps (See Supplementary Information). This increase in the variability of the reconstructions is needed to avoid an overconfident estimation of $r(n)$.

Performances for the two experiments are shown in Figs~\ref{Fig4}c and~\ref{Fig4}e. Grid scans reduce $r(n)$ linearly with increasing $n$. The decision algorithm outperforms a simple grid scan and is nearly optimal. When most of the current gradient is localised, the grid scan is far from optimal and even the decision algorithm has more room for improvement. In this case, the performance of the algorithm is determined by how representative the training data is. Quantitative analysis of all 10 examples is in the Supplementary Information.

We propose a simple stopping criterion that uses the estimated reduction of the error $r(n)$ to determine when to stop measuring a given current map, in a scenario where more experiments are waiting to be conducted. For a given current map, from which $n$ pixels have been measured, the error after the next measurement batch is estimated for reconstruction $m$ to be $\tilde{r}_m(n+\Delta)$, where $\Delta$ is the size of the batch.
Thus the estimated rate at which the error decreases is 
$\beta_m \equiv \bigl\lvert\tilde{r}_m(n+\Delta)-\tilde{r}_m(n)\rvert/\Delta$.
In the worst case among the candidate reconstructions, this rate is $\beta \equiv \min_m \beta_m$.
However, if the algorithm begins to measure a new map, for which no reconstructions yet exist, the error of that map will decrease at a rate of at least $\alpha \equiv 1/N$; this  is the slope achieved by a grid scan and the worst case of the decision algorithm (black lines in Fig \ref{Fig4}c,e).
Hence when $\beta<\alpha$, it is beneficial to halt measurement and move onto a new current map that is awaiting measurement. Since $\alpha$ and $\beta$ are the worst-case estimates for each case, the criterion is conservative.
The stopping points by this criterion are shown in Figs.~\ref{Fig4}(c,e) with orange dashed lines.
The total average time (measurement time plus decision time) to reach the stopping criterion was 237~s, compared with 554~s to measure the complete current map by grid scan, reducing the time needed by a factor between 1.84 and 3.70 across all 10 test cases. 
A more sophisticated stopping criterion utilising the number of remaining unmeasured current maps and total measurement budget is given in Methods.

\section*{Generality}
To prove the versatility of the algorithm, which does not require assumptions about the characteristics of the acquired data, we applied it to a different measurement configuration also encountered in quantum dot tuning. In this configuration the current flowing through the device is measured as a function of two gate voltages ($V_1 $ and $V_2$), while keeping other voltages fixed ($V_\text{G}$, $V_{\text{bias}}$, $V_3$ and $V_4$). The current map in this case has large areas where the current scarcely changes, and diagonal features indicative of Coulomb blockade. For the training set, we measured 382 current maps with a resolution of $251\times251$ which we cropped randomly with simple image augmentation techniques.

\begin{figure*}
\includegraphics[width=\textwidth]{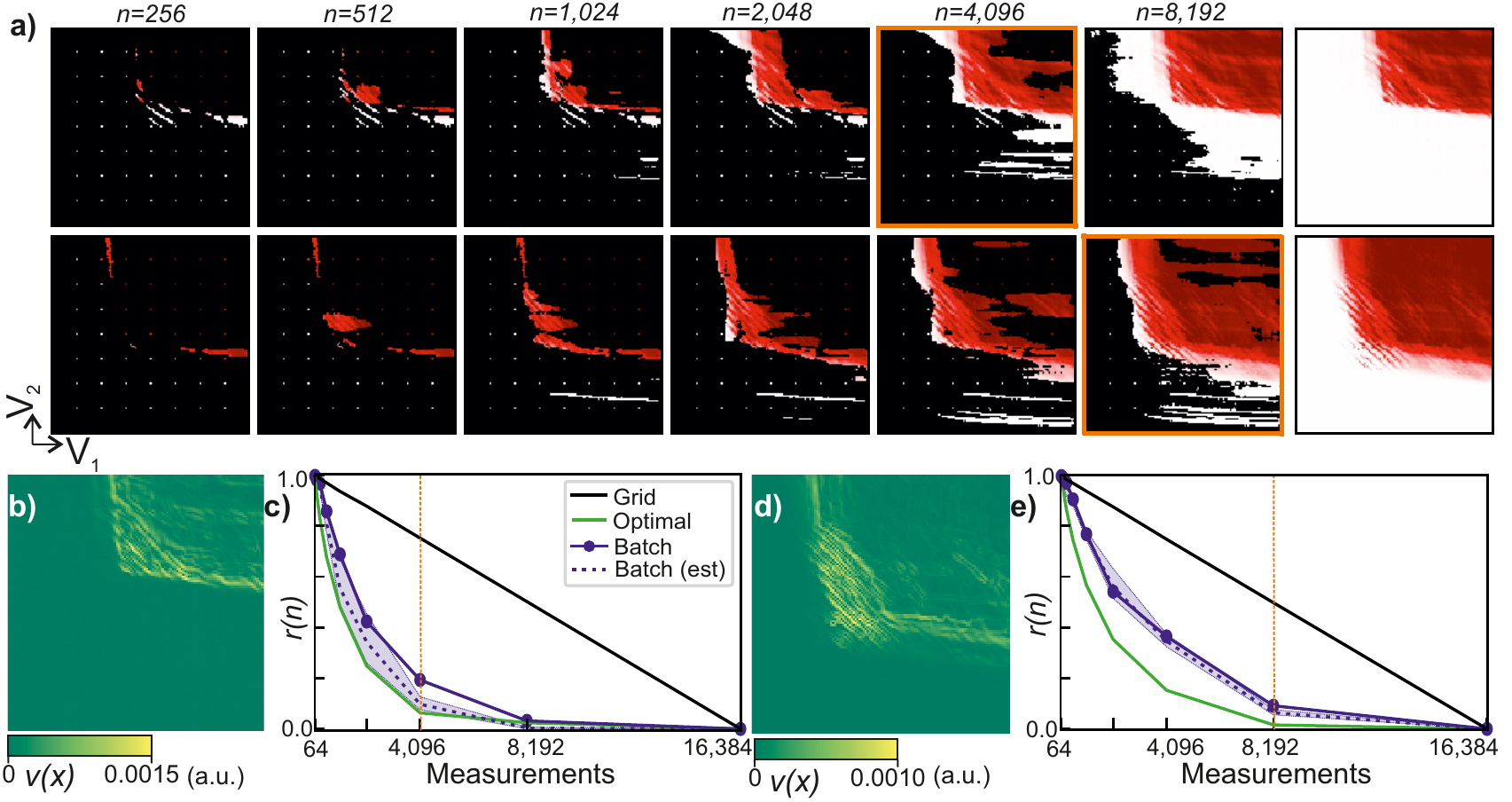}
\caption{\label{Fig5}
Algorithm-assisted measurements of a current map. (a) Sequential batch measurement. Each row displays the algorithm assisted measurements of a current map as a function of $V_1$ and $V_2$ for different values of $n$. The last plot in each row is the full-resolution current map. (b,d) Current gradient map for both examples in a. (c,e) Measure of the algorithm's performance $r(n)$, average real-time estimate of $r(n)$ with 90\% credible interval, and optimal $r(n)$ for both current maps in (a). The black line is the value of $r(n)$ corresponding to the alternating grid scan method. The dashed orange line indicates the value of $n$ determined by the stopping criterion. The alternating grid scan took 2,267~s and 2,333~s to acquire all measurements in the two cases. The batch method took 673~s and 1,552~s to reach the stopping criterion.
}
\end{figure*}
We tested the performance of the algorithm in this new scenario by taking two different combinations of $V_\text{G}$, $V_{\text{bias}}$, $V_3$ and $V_4$ and measuring the corresponding current maps in real time (Fig.~\ref{Fig5}). The device was thermally cycled after the training set was acquired and also between the acquisition of the two current maps in Fig.~\ref{Fig5}. The algorithm focused on measuring regions of high current gradient, the corner edges and, in particular, the Coulomb peaks close to these.  

In the top row of Fig.~\ref{Fig5}a, $n=4,096$ was chosen by the stopping criterion. In the bottom row, the corners edges extended further in the current map and the stopping criterion chose $n=8,192$. This reduced the time needed to measure the current maps by 3.36 and 1.50 for the two test cases when compared with the alternating grid scans. 

\section*{Automating what to measure next}
The proposed measurement algorithm makes real-time informed decisions on which measurements to perform next on a single quantum dot device. Decisions are based on the disagreement of competing reconstructions built from sparse measurements. The algorithm outperforms grid scan in all cases, and in the majority of cases shows nearly optimum performance. The algorithm reduced the time required to observe the regions of finite current gradient by factors ranging from 1.5 to 3.7 times.

Our algorithm  with no modifications can be re-trained to measure different current map configurations. It simply requires a diverse dataset of training examples from which to learn. The decision algorithm performed well even when trained on a small data set of only 382 current maps (at a resolution of $251\times251$), implying that it is robust to limited training datasets. Our algorithm focused on observing all informative regions present in the current map. However, if only specific features such as Coulomb peaks or Coulomb diamond edges are of interest, the acquisition function can be specifically designed to focus on them (see Supplementary Information).
\newline
We believe that our algorithm represents a significant first step in automating what to measure next in quantum devices. For a single quantum dot it provides a means of accelerating what can currently be achieved by human experimenters and other automation methods. Our algorithm can be applied to acquire data in other types of experiments provided an appropriate training data set is accessible and the generative model is retrained. It will not be long before this kind of approach enables experiments to be performed, and technology to be developed, that would not be feasible otherwise.

\section*{Methods}

\subsection*{Distribution of reconstructions and sampling}
Since it is known that deep generative models work well when the data range is from -1 to 1, all measurements are rescaled so that the maximum value of the absolute value of the initial measurement is 1. Let $Y$ be a random vector containing all pixel values. The CVAE model makes a distribution $p(Y \mid Y_i)$ and enables sampling from the distribution. Observation $Y_n$, where $n\geq 1$, is the set of pairs of location $x_j$ and measurement $y_j$: $Y_n=\{(x_j,y_j) \mid j=1,\ldots,n\}$. Also, a subset of measurements is defined: $Y_{n:n'}=\{(x_j,y_j) \mid j=n,\ldots,n'\}$. The likelihood of observations given $Y$ is defined by
\begin{equation}
p(Y_n \mid Y) \propto \exp \bigl(-\lambda\Sigma_{(x,y)\in Y_n} \lvert y-Y(x)\rvert \bigr) \, ,
\label{eqn:L}
\end{equation}
where $Y(x)$ is the pixel value of $Y$ at $x$, and $\lambda$ is a free parameter that determines the sensitivity to the distance metric and is set to 1.0 for all experiments in this paper. The posterior probability distribution is defined by Bayes' rule:
\begin{equation}
p(Y\mid Y_n) \propto p(Y_n\mid Y)\,p(Y) \, .
\label{eqn:PosteiorY}
\end{equation}
Likewise, we can find the posterior distribution of $z$ given measurements instead of $Y$. Let $z'$ denote another input of the decoder, which is set to $Y_i$ in the experiments. Then the posterior distribution of $z$ can be expressed with $z'$ when $n \geq i$:
\begin{align*}
p(z\mid Y_n,z') & \propto p(z\mid z')\, p(Y_n\mid z,z') \\
& \propto p(z) \int_Y p(Y_n\mid Y)\, p(Y\mid z,z')\, dY \\
& \propto p(z)\, p(Y_n\mid Y=\hat{Y}_z) \, ,
\end{align*}
where $\hat{Y}_z$ is the reconstruction produced by the decoder given $z$ and $Y_i$. Since all inputs of the decoder are given, $p(Y\mid z,z')$ is the Dirac delta function centered at $\hat{Y}_z$. Also, $p(z\mid z')=p(z)$ as $z$ and $z'$ are independent. Proposal distribution for MH is set to a multivariate normal distribution having centered mean and a covariance matrix equal to one quarter of the identity matrix. For the experiments in this paper, 400 iterations of MCMC steps are conducted when $n=64\times 2^{b}$, where $b$ is any integer larger than or equal to 0. We found that 400 iterations result in good posterior samples. If $(x_{n+1},y_{n+1})$ is newly observed, then the posterior can be updated incrementally: 
\begin{align*}
p(z\mid Y_{n+1},z')&=\frac{p(x_{n+1},y_{y+1}\mid z,z')}{p(x_{n+1},y_{n+1}\mid Y_n,z')}\, p(z\mid Y_n,z') \\
&=\frac{p(x_{n+1},y_{y+1}\mid \hat{Y}_z)}{p(x_{n+1},y_{n+1}\mid Y_n,z')}\, p(z\mid Y_n,z') \, .
\end{align*}
because each term in (\ref{eqn:L}) can be separated. 

\subsection*{Decision algorithm}
In this section, we derive a computationally simple form of the information gain and the fact that maximising the information gain is equal to minimising the entropy. Let $p_n(\cdot) = p(\cdot|Y_n,z')$, and any probabilistic quantity of $y_{n+1}$ has the condition $x_{n+1}$, but omitted for brevity.

The continuous version of the information gain equation is
\begin{align}
&\mathbb{E}_{y_{n+1}}\Bigl[ \textrm{KL}\bigl( p_n(z\mid y_{n+1}) \| p_n(z) \bigr) \Bigr] \nonumber \\
&=\int_{y_{n+1}} p_n(y_{n+1}) \textrm{KL}\bigl( p_n(z\mid y_{n+1}) \| p_n(z) \bigr) dy_{n+1} \nonumber \\
&=\int_{y_{n+1}} p_n(y_{n+1}) \int_z p_n(z\mid y_{n+1}) \log \frac{p_n(z\mid y_{n+1})}{p_n(z)}  dzdy_{n+1} \label{eqn:interm} \\
&=\int_{y_{n+1}}\int_z p_n(z,y_{n+1}) \log \frac{p_n(z,y_{n+1})}{p_n(z)p_n(y_{n+1})}dzdy_{n+1} \nonumber \\
&=I(z\mid Y_n \,;\, y_{n+1}\mid Y_n) \, , \nonumber
\end{align}
where KL is Kullback-Leibler divergence, $I(\cdot;\cdot)$ is mutual information. 
Since $I(z\mid Y_n \,;\, y_{n+1}\mid Y_n)=\mathrm{H}(z\mid Y_n) - \mathrm{H}(z\mid Y_n,y_{n+1})$, maximising the expected KL divergence is equivalent to minimising $\mathrm{H}(z\mid Y_n,y_{n+1})$, which is the entropy of $z$ after observing $y_{n+1}$.

Since this integral is hard to compute, we approximate probability density functions (PDFs) with samples and substitute them into (\ref{eqn:interm}). Let $n_s$ is the number of measurements that are used for sampling reconstructions $\hat{z}_1,\ldots,\hat{z}_M$ (the samples can be converted to $\hat{Y}_1,\ldots,\hat{Y}_M$). Then $p_{n_s}(z) \approx \frac{1}{M}\sum_m\delta_{\hat{z}_m}(z)$, or with the sample index $m$, $P_{n_s}(m)=1/M$. For any $n\geq n_s$, the probability is updated with the new measurements after $n_s$: $P_n(m;n_s)=\frac{p(Y_{n_s+1:n}|\hat{Y}_m)}{\Sigma_m p(Y_{n_s+1:n}|\hat{Y}_m)}$. For brevity, the sampling distribution information $n_s$ is omitted for the remaining section. Likewise, $p_n(y_{n+1})=\int_z p_n(y_{n+1}\mid z)\, p_n(z)\approx \sum_m P_n(m)\, p_n(y_{n+1}\mid z_m)$. Lastly, we use the value of $\hat{Y}_m$ at $x_{n+1}$ for a sample of $p_n(y_{n+1}\mid z_m)$ for simple and efficient computation. As a result, the information gain is approximated by:
\begin{align*}
&\mathbb{E}_{y_{n+1}}\Big[ \textrm{KL}\bigl( p_n(z\mid y_{n+1}) \,\|\, p_n(z) \bigr) \Big] \\
&\approx \sum_m P_n(m) \, \textrm{KL}(P_{n+1} \,\|\, P_n) \, .
\end{align*}

\subsection*{Simulator for Training data}
To aid the training of the model simulated training data was used to prevent over-fitting. Simulated data produced via a simple implementation of the constant interaction model~\cite{Hanson2007} was used along with basic data augmentation techniques. These techniques were not intended to be physically accurate but instead to produce quickly a diverse set of examples that contain features that mimic real data.

The constant interaction model makes the assumptions that all interactions felt by a confined electrons within the dot can be captured by a simple constant capacitance $C_\Sigma$ which is given by $C_\Sigma=C_\text{S}+C_\text{D}+C_\text{G}$ where $C_\text{S}$, $C_\text{D}$ and $C_\text{G}$ are capacitances to the source, drain and gate respectively. Making this assumption the total energy of the dot $U(N)$ where $N$ is the number of electrons occupying the dot, is $U(N)=\frac{(-|e|(N-N_0)+C_\text{S}V_\text{S}+C_\text{D}V_\text{D}+C_\text{G}V_\text{G})^2}{2C_\Sigma}+\sum\limits_{n=1}^N E_n$ where $N_0$ compensates for the background charge and $E_n$ is a term that represents occupied single electron energy levels that is characterised by the confinement potential. 

Using this we derive the electrochemical potential $\mu(N)=U(N)-U(N-1)=\frac{e^2}{C_\Sigma}(N-N_0-\frac{1}{2})-\frac{|e|}{C_\Sigma}(V_\text{S}C_\text{S}+V_\text{D}C_\text{D}+V_\text{G}C_\text{G})+E_n$.

To produce a training example random values are generated for $C_\text{S}$, $C_\text{D}$ and $C_\text{G}$. The energy levels within a randomly generated gate voltage window and source drain bias window are then counted. To aid generalisation to real data we randomly generated energy level transitions (which are also counted) as well as slightly linearly scaled $C_\Sigma$, $C_\text{S}$, $C_\text{D}$, and $C_\text{G}$ with $N$. This linear scaling was also randomly generated and results in produced diamonds that vary in size with respect to $V_\text{G}$. Examples of the training data produced by this simulator can be seen in Supplementary Material.

\subsection*{Stopping criterion}
Utility, denoted by $u$, is the ratio of total measured gradient to the total gradient of a stability diagram: $u(n)=1.0-r(n)$. Here, we assume that we have $K$ more stability diagrams to be measured. The location of each diagram is defined by a different voltage range, and $k=0,\ldots,K$ is the index of the diagrams, where $k=0$ is the index of the diagram that we are currently measuring.

Let $T$ denote the total measurement budget for the current and remaining stability diagram. In this paper we assume that a unit budget for measuring one pixel is 1.0. The total utility is
\begin{align*}
u_\text{tot}&=\sum_{k=0}^K u_k(t_k) \\
&=u_0(t_0) + u_\text{nxt}(T-t_0) \,,
\end{align*}
where $u_k(\cdot)$ is the utility from measuring $k$th diagram, $t_k$ is the planned budget for $k$th diagram satisfying $\sum_{k=0}^Kt_k=T$, and $u_\text{nxt}(T-t_0)=\sum_{k=1}^Ku_k(t_k)$. 

Let $t$ denote the already spent budget on the current diagram, $t\leq t_0$. If we stop the measurement then $t_0=t$, or $t_0=t+\Delta$ if we decide to continue the measurement, where $\Delta$ is a predefined batch size.
For the decision, the utilities of two cases are compared: when $t_0=t$,
\begin{equation}
u_\text{tot}=u_0(t) + u_\text{nxt}(T-t) \, .
\label{eqn:move}
\end{equation}
Otherwise, $t_0=t+\Delta$ and 
\begin{equation}
u_\text{tot}=u_0(t+\Delta) + u_\text{nxt}\bigl( T-(t+\Delta) \bigr) \quad .
\label{eqn:stay}
\end{equation}
If (\ref{eqn:stay}) $<$ (\ref{eqn:move}), it is better to stop and move to the next voltage range. Rearranging the inequality leads to
\begin{equation}
u_0(t+\Delta)-u_0(t) 
< u_\text{nxt}(T-t) - u_\text{nxt} \bigl( T-(t+\Delta) \bigr) \, .
\label{eqn:ineq}
\end{equation}
The left-hand-side (lhs) of ($\ref{eqn:ineq}$) means the difference of utility if we invest $\Delta$ budget more on the current diagram, and the right-hand-side the difference when $\Delta$ more budget is used for remaining diagrams. As we discussed in Results section, we can calculate multiple slope estimates $\beta_m$ for spending $\Delta$ to the current diagram: $ u_0(t+\Delta)-u_0(t) \approx  \beta_m \Delta$.

The right-hand-side (rhs) of (\ref{eqn:ineq}) can be approximated by $\alpha\Delta$ if $K=\infty$, where $\alpha=1/16,384$ is the slope of grid scan measuring a new stability diagram. Note that $\alpha$ can be considered as the empirical worst case performance of the decision algorithm measuring a new diagram as it holds for all the experiments we have conducted. If $\Delta=N$, this approximation is the exact quantity for any algorithms as all algorithms satisfy $r(0)=1.0$ and $r(N)=0.0$. Since $\alpha$ can be interpreted as the worst case estimate, we also approximate lhs of (\ref{eqn:ineq}) with the worst case estimate $\beta=\min_m \beta_m$.

If $K<\infty$, and the remaining budget $T-t$ is more than the budget to measure all of remaining diagrams, there is no utility after all measurements are finished. Hence, the approximation is capped:
$$
u_\text{nxt}(T-t)=\alpha \min (T-t, N \times K) \, ,
$$
where $K$ is the number of remaining diagrams to be measured.

As a result, the stopping criterion when $K=\infty$ is
$$
\beta < \alpha \, .
$$
The stopping criterion when $K<\infty$ is
\begin{equation}
\beta < \frac{\alpha (\min(T-t,N\times K) - \min(T-(t+\Delta),N\times K))}{\Delta} .
\label{eqn:stop_general}
\end{equation}
The rhs of (\ref{eqn:stop_general}) is always less than or equal to $\alpha$, and more total budget $T$ makes it low, which leads to late stopping or no stopping.

\section*{addendum}
\subsection*{Acknowledgements}
We acknowledge discussions with J.A. Mol and S.C. Benjamin.
This work was supported by the EPSRC National Quantum Technology Hub in Networked Quantum Information Technology (EP/M013243/1), Quantum Technology Capital (EP/N014995/1), Nokia, Lockheed Martin, the Swiss NSF Project 179024, the Swiss Nanoscience Institute and the NCCR QSIT. This publication was also made possible through support from Templeton World Charity Foundation and John Templeton Foundation. The opinions expressed in this publication are those of the authors and do not necessarily reflect the views of the Templeton Foundations. We acknowledge J. Zimmerman and A. C. Gossard for the growth of the AlGaAs/GaAs heterostructure.

\subsection*{Author Contributions}
D.T.L. and N.A. and the machine performed the experiment. H.M. developed the algorithm under the supervision of M.A.O. The sample was fabricated by L.C.C., L.Y., and D.M.Z. The project was conceived by G.A.D.B., E.A.L., M.A.O., and N.A. All authors contributed to the manuscript and commented and discussed results.  
\subsection*{Competing Interests}
The authors declare that they have no
competing financial interests.
\subsection*{Correspondence}
Correspondence and requests for materials
should be addressed to Natalia Ares~(email: natalia.ares@materials.ox.ac.uk).

\onecolumngrid

\vfill

\newpage

\appendix
\newcommand{\beginsupplement}{%
        \setcounter{table}{0}
        \renewcommand{\thetable}{S\arabic{table}}%
        \setcounter{figure}{0}
        \renewcommand{\thefigure}{S\arabic{figure}}%
     }
     
\beginsupplement

\section{Supplementary}

\subsection*{Training and loss function}
The purpose of this section is to define the losses used for the training of the CVAE. The training is performed by minimising user-defined loss terms through changing the decoder and encoder parameters $\theta$ and $\phi$ using a gradient decent based method. The two loss terms that are minimised to train the encoder and decoder networks are the difference loss and the latent loss. 

The difference loss consists of two difference metrics. The first is a sum of the pixel-wise difference between the reconstruction and the training example. The second is a contextual difference which is similar in concept to GAN; the contextual loss is taken from another convolutional neural network called the discriminator. The discriminator is trained in tandem with the encoder and decoder and is trained to distinguish between reconstructions and training examples. The input to the discriminator is a training example $Y$ or reconstruction $\hat{Y}$ and the output is a value between 0 and 1, representing the probability the input is a training example or a reconstruction. As the discriminator is trained to distinguish between training examples and reconstructions, it learns to decode contextual features that distinguish reconstructions from training examples. We then calculate the difference between intermediate layer representations of the training example and intermediate layer representations of its reconstruction. If we ignore the contextual loss, the decoder produces only blurry reconstructions. 

The latent loss is applied only to the encoder and forces the set of encoded training examples $\{z \}$ to be normally distributed with mean of the zero vector and the covariance of a diagonal matrix. This can be achieved by minimising the Kullback-Leibler (KL) divergence between the output distribution of the encoder and the target zero-mean distribution.

\subsection*{Network specification}
The specification of the convolutional neural networks used in this paper is described in Table~\ref{method:net_encoder}$\sim$\ref{method:net_discriminator}. Exponential linear unit is applied after each layer except the final layer of the encoder, decoder, and the discriminator. Batch normalisation is applied after all convolution layers except as separately described. The first and second number in parentheses of the layer names indicate kernel size and stride.

\subsection*{Noisy reconstruction}
For the estimation $\tilde{r}_m(n)$, a single reconstruction $\hat{Y}_m$ is augmented to 30 noisy reconstructions:
$$
\hat{Y}_{m,j,\text{SNR}}(x) = \hat{Y}_m(x) + \alpha_{m,j,\text{SNR}} \times \mathcal{E}_j(x) \quad \text{for all} \quad x \,\,\text{in}\,\, X ,
$$
where $\mathcal{E}_j = \{(x, \epsilon^j_x)|x\in X\}$ is a noise profile consisting of pairs of location and noise, $X$ is a set of all voltage pairs in a 2D domain, and $\alpha_{m,j,\text{SNR}}$ is a multiplier that makes the signal-to-noise ratio $\text{SNR}$, where the signal is $\hat{Y}_m$ and the noise is $\mathcal{E}_j$. We measured 10 noise profiles at non-conducting voltage ranges, but very close to Coulomb diamonds, and $j$ is the index of the profile. $\text{SNR}$ is chosen from $\{20^2,40^2,80^2\}$, which leads to a high noise, medium noise, and low noise.

\begin{figure}
\centering
\includegraphics[width=\textwidth]{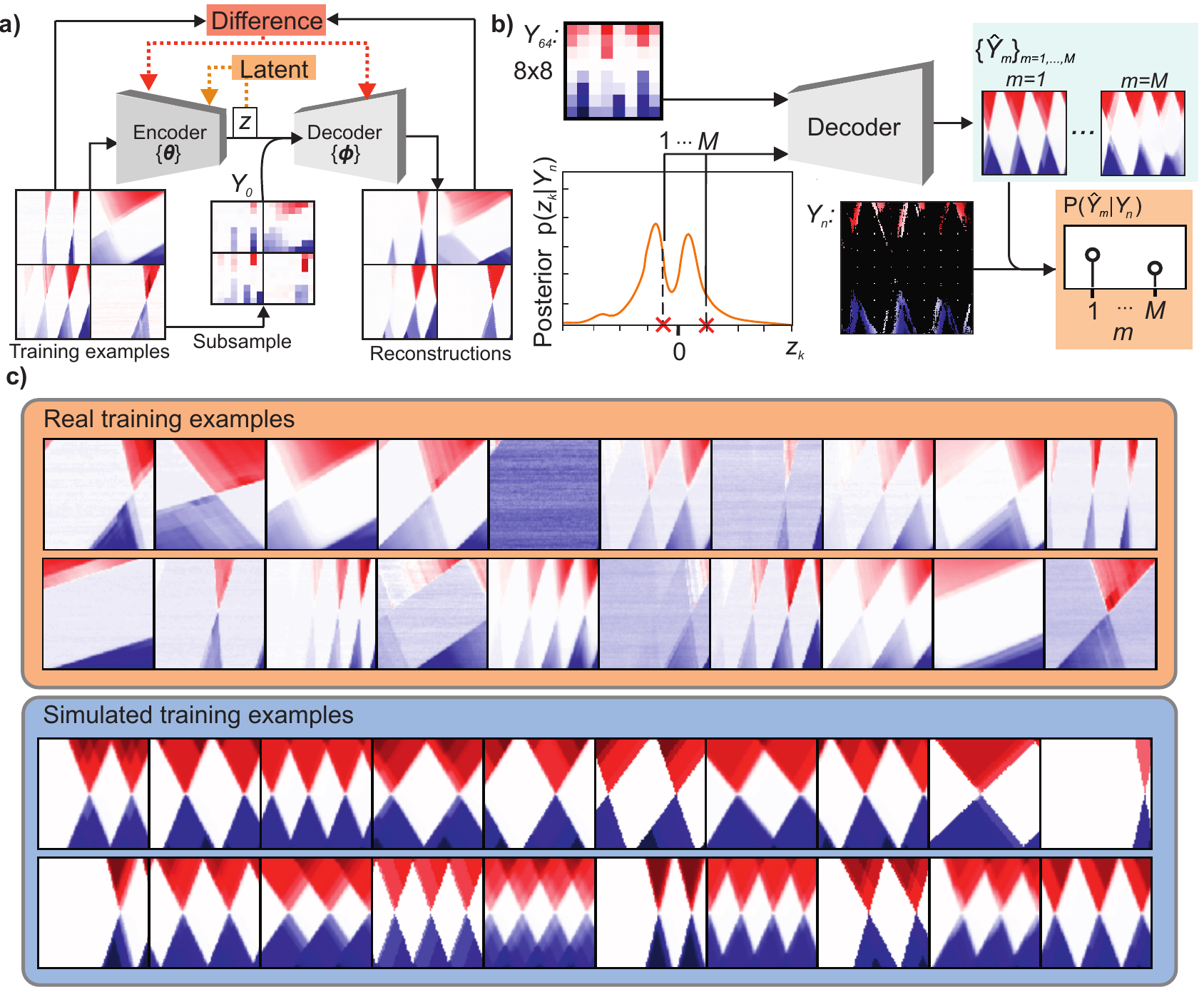}
\caption{\label{sup1}(a) Training procedure. Training examples are converted to latent vectors by the encoder. Latent vectors and 8$\times$8 sub-sampled training examples are transformed into reconstructions by the decoder. The difference (red box) between original examples and reconstructions is used to optimise the encoder/decoder parameters $\theta$ and $\phi$. The distribution of training examples in the latent space is enforced during training by latent loss (orange box).
(b) Generation of reconstructions. After 8$\times$8 initial measurements, latent vectors are sampled from the posterior distribution of $z$ and transformed by the decoder to generate multiple reconstructions $\hat{Y}_1,\ldots,\hat{Y}_M$. Posterior probability for reconstructions $P(\hat{Y}_m|Y_{n})$ is calculated with respect to acquired partial measurements $Y_{n}$. (c) Real and simulated training examples.}
\end{figure}

\newpage
\begin{table}
\centering
\begin{tabular} {l r} 
Layer name & output size \\
\hline
Initial & 128x128x1 \\
Conv(5,2) & 64x64x64 \\
Max pooling(3,2) & 32x32x64 \\

Conv(3,1) & 32x32x128 \\
Conv(3,2) & 16x16x128 \\

Conv(3,1) & 16x16x128 \\
Conv(3,2) & 8x8x128  \\

Conv(3,1) & 8x8x128  \\
Conv(3,2) & 4x4x128  \\
Fully connected & 200 \\

\end{tabular}
\caption{\label{method:net_encoder}
Specification of the encoder. }
\end{table}

\begin{table}
\centering
\begin{tabular} {l r} 
Layer name & output size \\
\hline
Initial & 1x1x(100+64) \\

Conv'(3,2) & 2x2x1,024 \\
Conv(3,1) & 2x2x1,024 \\

Conv'(3,2) & 4x4x512 \\
Conv(3,1) & 4x4x512 \\

Conv'(3,2) & 8x8x256 \\
Conv(3,1) & 8x8x256 \\

Conv'(3,2) & 16x16x128 \\
Conv(3,1) & 16x16x128 \\

Conv'(3,2) & 32x32x64 \\
Conv(3,1) & 32x32x64 \\

Conv'(3,2) & 64x64x64 \\
Conv(3,1) & 64x64x64 \\

Conv'(3,2) & 128x128x32 \\
Conv(3,1) & 128x128x32 \\

Conv(1,1,tanh) & 128x128x1 \\

\end{tabular}
\caption{\label{method:net_decoder}
Specification of the decoder. }
\end{table}

\begin{table}
\centering
\begin{tabular} {l r r} 
Layer name & output size & remark \\
\hline
Initial & 128x128x1 & \\
Conv(5,2) & 64x64x64 & context loss\\

Conv(3,1) & 64x64x128 & No BN\\
Conv(3,2) & 32x32x128 & context loss\\

Conv(3,1) & 32x32x128 & No BN\\
Conv(3,2) & 16x16x128 & context loss\\

Conv(3,1) & 16x16x128 & No BN\\
Conv(3,2) & 8x8x128  & context loss\\

Conv(3,1) & 8x8x128  & No BN\\
Conv(3,2) & 4x4x128  & context loss\\
Global average pooling & 1x1x128 & context loss \\
Fully connected & 2 & \\

\end{tabular}
\caption{\label{method:net_discriminator}
Specification of the discriminator. In remarks, \textit{NO BN} indicates that batch normalisation is not applied to the layer, and \textit{context loss} indicates that the layer is used to calculate the context loss.}
\end{table}

\newpage

\begin{figure}
\centering
\includegraphics[width=\textwidth]{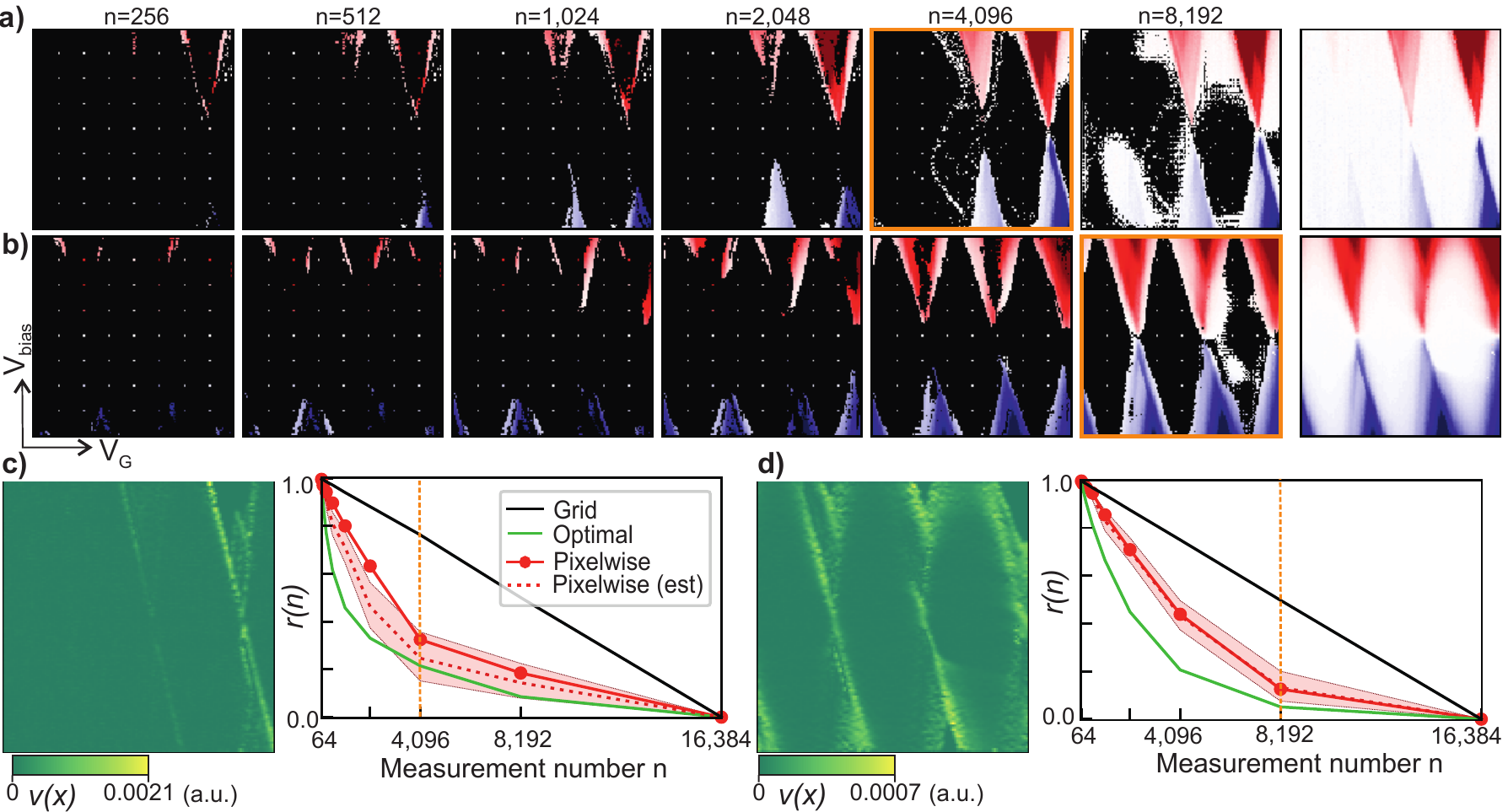}
\caption{\label{aqu2} Intermediate steps and quantitative analysis of the pixel-wise decision method}
\end{figure}

\begin{figure}
\centering
\includegraphics{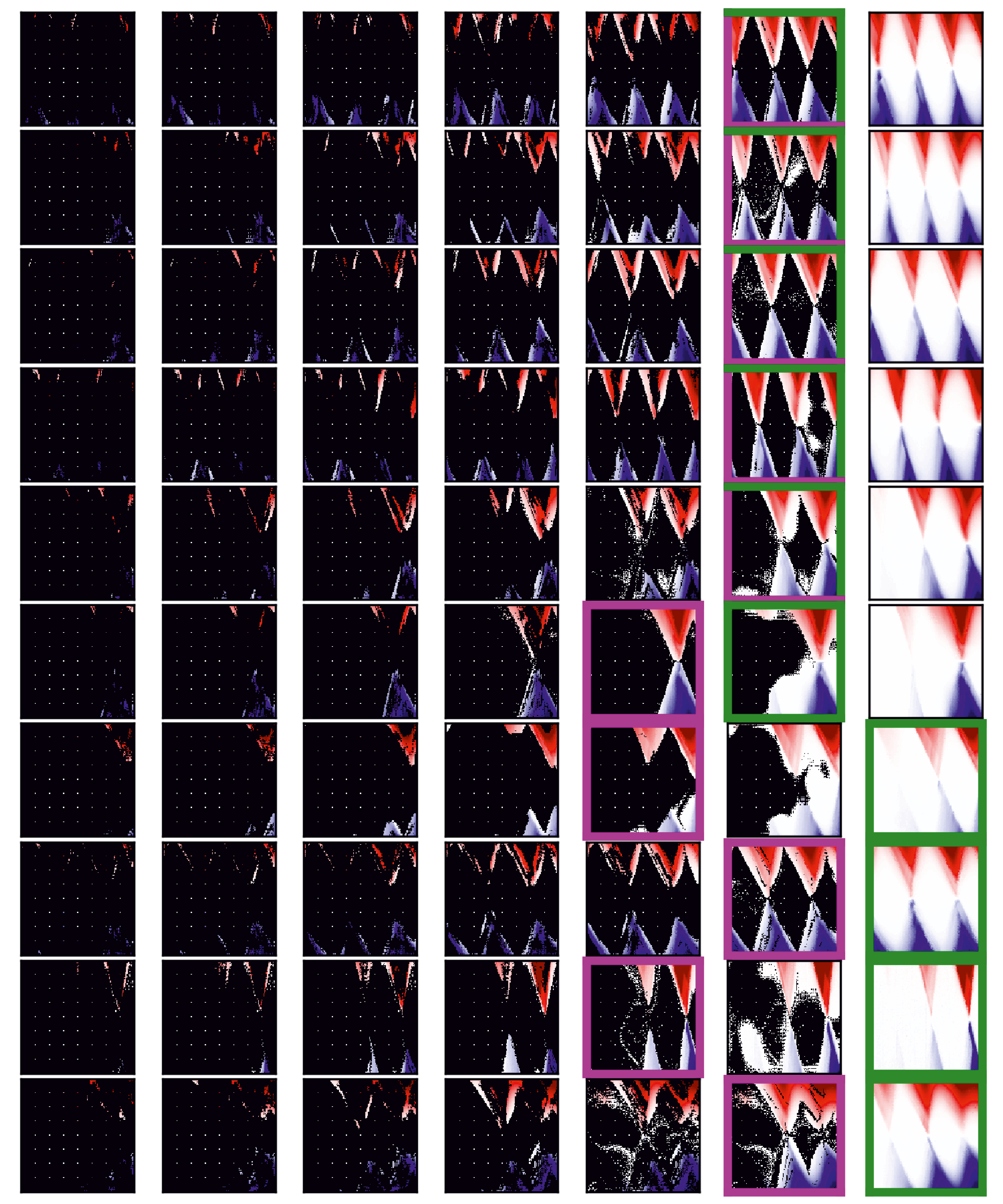}
\caption{\label{supple:decision_pointwise}
Each row shows intermediate steps of point-wise decision for given voltage ranges. Magenta box indicates the default stopping criterion, and green box indicates when we have allocated a measurement budget of 70\% for full measurement of all 10 examples.}
\end{figure}

\begin{figure}
\centering
\includegraphics{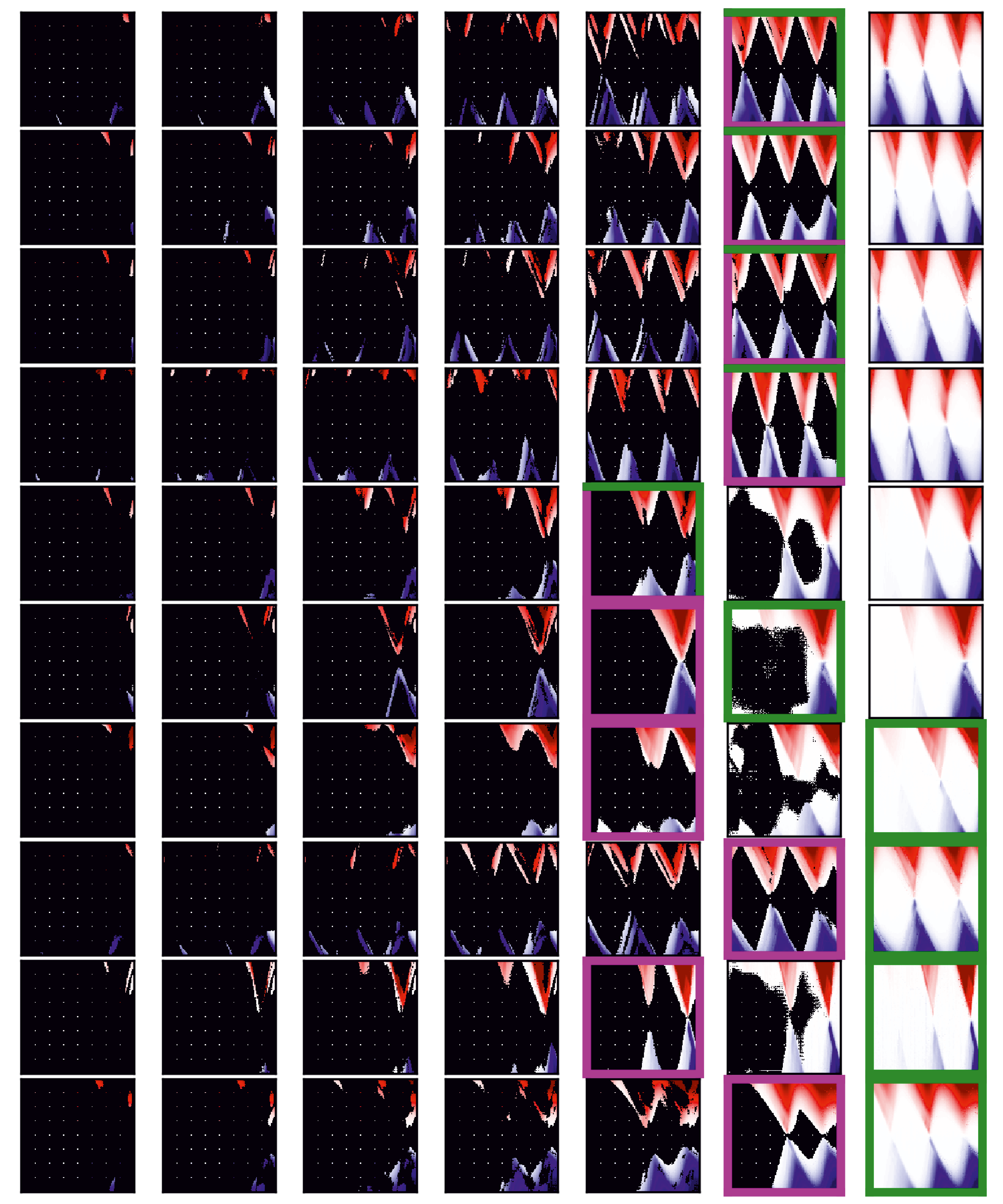}
\caption{\label{supple:decision_batch}
Each row shows intermediate steps of batch decision for given voltage ranges. Magenta box indicates the default stopping criterion, and green box indicates when we have 70\% budget for full measurement of all 10 examples.}
\end{figure}

\begin{figure}
\centering
\includegraphics{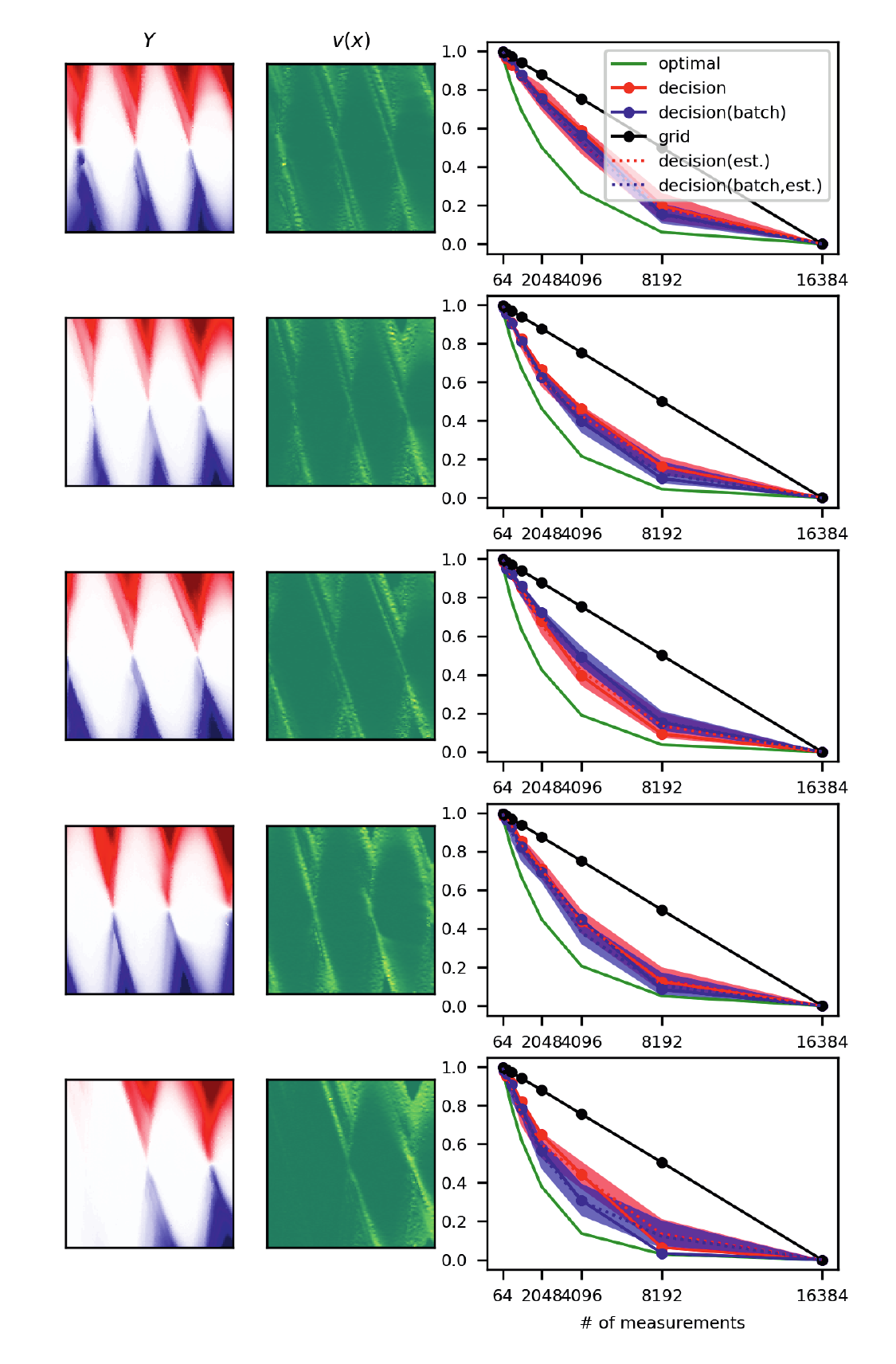}
\caption{\label{supple:result_all1}
Quantitative analysis for experiment number 1$\sim$5.}
\end{figure}

\begin{figure}
\centering
\includegraphics{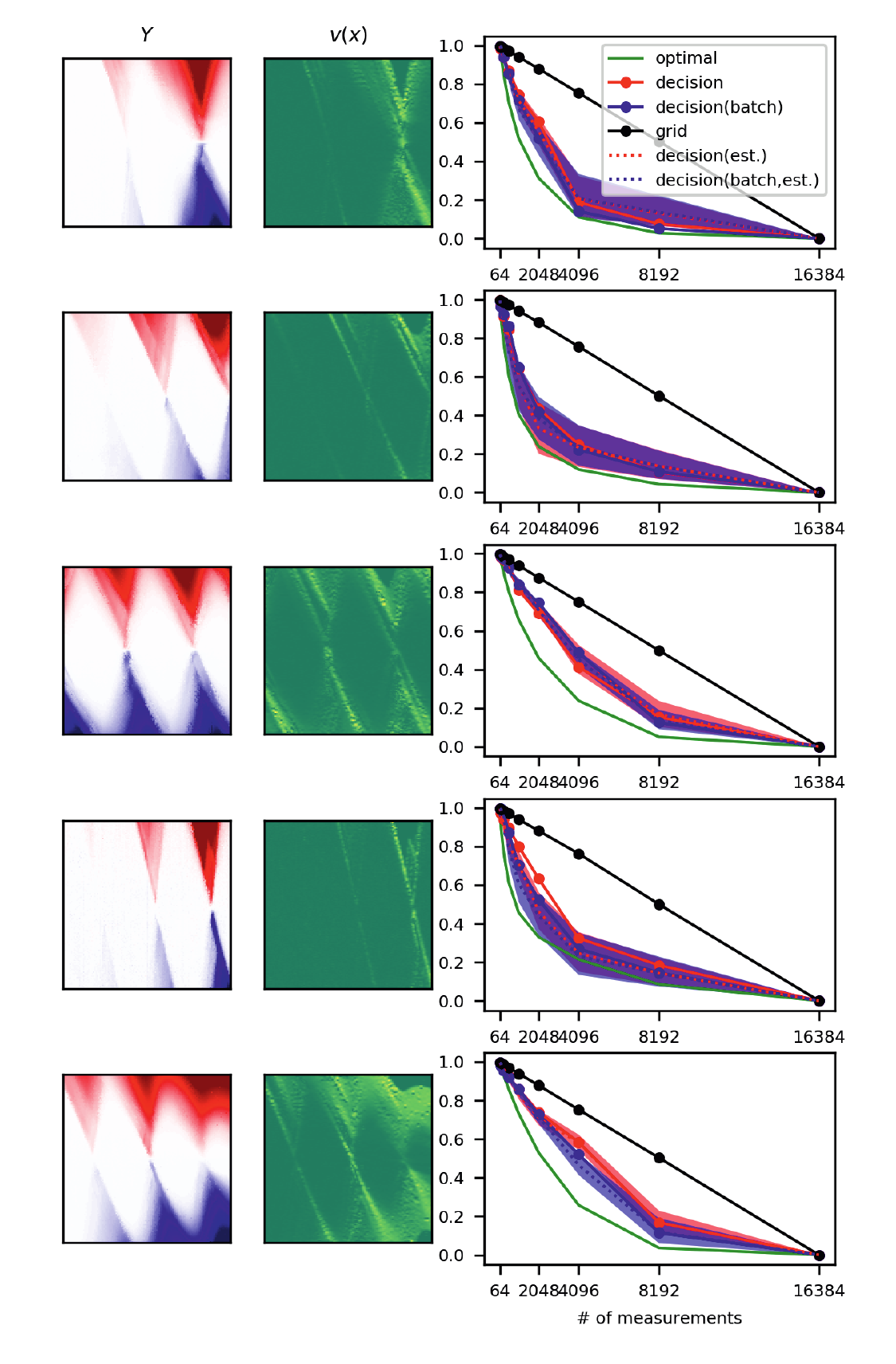}
\caption{\label{supple:result_all2}
Quantitative analysis for experiment number 6$\sim$10}
\end{figure}

\newpage

\begin{table}
\centering
\begin{tabular} {l | r r r r r r r r r} 
case index & 64 & 128 & 256 & 512 & 1,024 & 2,048 & 4,096 & 8,192 & 16,384 \\
\hline
1&10.50&13.13&18.32&28.07&47.27&84.51&156.98&294.91&561.21 \\
2&10.14&12.71&17.46&26.78&44.95&81.32&152.31&289.03&554.98 \\
3&10.49&13.08&17.77&26.66&45.97&82.82&53.86&291.15&557.51 \\
4&9.40&12.16&17.09&27.34&46.69&82.99&152.64&289.00&555.27 \\
5&10.52&13.12&18.10&26.99&45.66&81.28&151.10&287.97&553.07 \\
6&10.14&12.73&17.21&27.07&45.17&80.55&149.18&285.82&550.95 \\
7&15.71&18.34&22.77&31.73&50.61&87.45&159.77&296.92&563.10 \\
8&14.96&17.85&22.53&32.66&52.08&89.56&161.62&300.80&566.80 \\
9&10.14&12.82&17.57&26.89&45.08&80.56&149.62&286.04&551.51 \\
10&15.31&17.88&22.71&31.71&51.10&87.94&159.44&296.00&561.27 \\
\end{tabular}
\caption{\label{supple:time_table1}
Measurement time for the batch method}
\end{table}

\begin{table}
\begin{tabular} {l | r r r r r r r r r} 
case index & 64 & 128 & 256 & 512 & 1,024 & 2,048 & 4,096 & 8,192 & 16,384 \\
\hline
1&10.51&13.53&18.52&28.27&46.09&81.51&149.73&285.62&552.69 \\
2&10.12&13.13&18.09&27.73&45.50&80.93&149.20&285.43&552.76 \\
3&10.50&13.26&18.19&27.78&45.66&80.98&149.22&282.37&546.65 \\
4&9.38&12.16&17.09&26.75&44.49&80.07&148.31&284.37&551.89 \\
5&10.51&13.28&18.21&27.86&45.69&81.02&149.20&285.02&552.15 \\
6&10.15&12.93&17.87&27.51&45.41&80.80&148.97&285.47&552.62 \\
7&15.70&18.64&24.02&34.52&52.28&87.49&155.75&292.05&559.88 \\
8&14.97&17.89&23.25&33.73&51.51&86.83&155.02&288.72&553.71 \\
9&10.13&12.89&17.87&27.57&45.42&80.97&148.89&285.08&553.30 \\
10&15.34&18.28&23.64&34.15&52.05&87.50&155.68&291.94&559.65 \\
\end{tabular}
\caption{\label{supple:time_table2}
Measurement time for grid scanning}
\end{table}

\clearpage

\subsection*{Context-aware decision for stability diagrams}
By converting reconstructions to some context maps, we can make a decision related with the context map. We have developed a segmentation method, that produces a segmentation map which has a value is 1 if the location is inside a diamond or 0 otherwise. This segmentation method is based on another deep neural network called a U-net~\cite{unet}. Training data for the segmentation network are pairs of current map and segmentation map, which are generated by the same simulator used for the reconstruction network.  Fig.~S7 shows the segmentation result of a trained network for 10 real stability diagrams.

\begin{figure}
\centering
\includegraphics{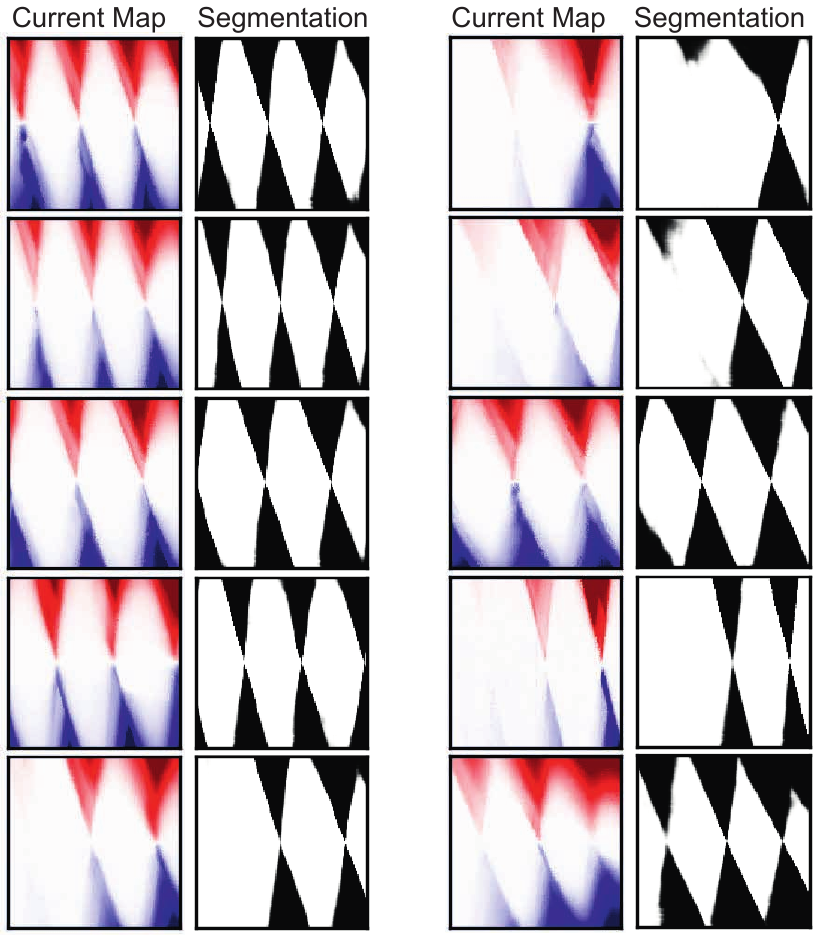}
\caption{\label{fig:segmentation}
Full resolution current map and segmented result from the segmentation network.
}
\end{figure}

By producing segmentation maps of reconstructions, their segmentation  disagreement can be calculated. This produces large disagreement along the edges of reconstructions resulting in measurements that focus on diamond edges as show in Fig.~S8a. Noise is also added to the outside of diamond segmented maps. This supplies further disagreement between segmentation maps which prioritises measurement outside of the diamond after edges are measured.

The success measure $e(n)$ in Fig.~S8b is calculated by applying the segmentation model to the fully measured current map and then applying a Sobel filter to the resulting segmented map; this produces an edge map. The error and optimal performance are then calculated as the ratio of this remaining quantity in the same way as was done for $r(n)$ except substituting the edge maps for transconductance maps.

\begin{figure}
\centering
\includegraphics[width=\textwidth]{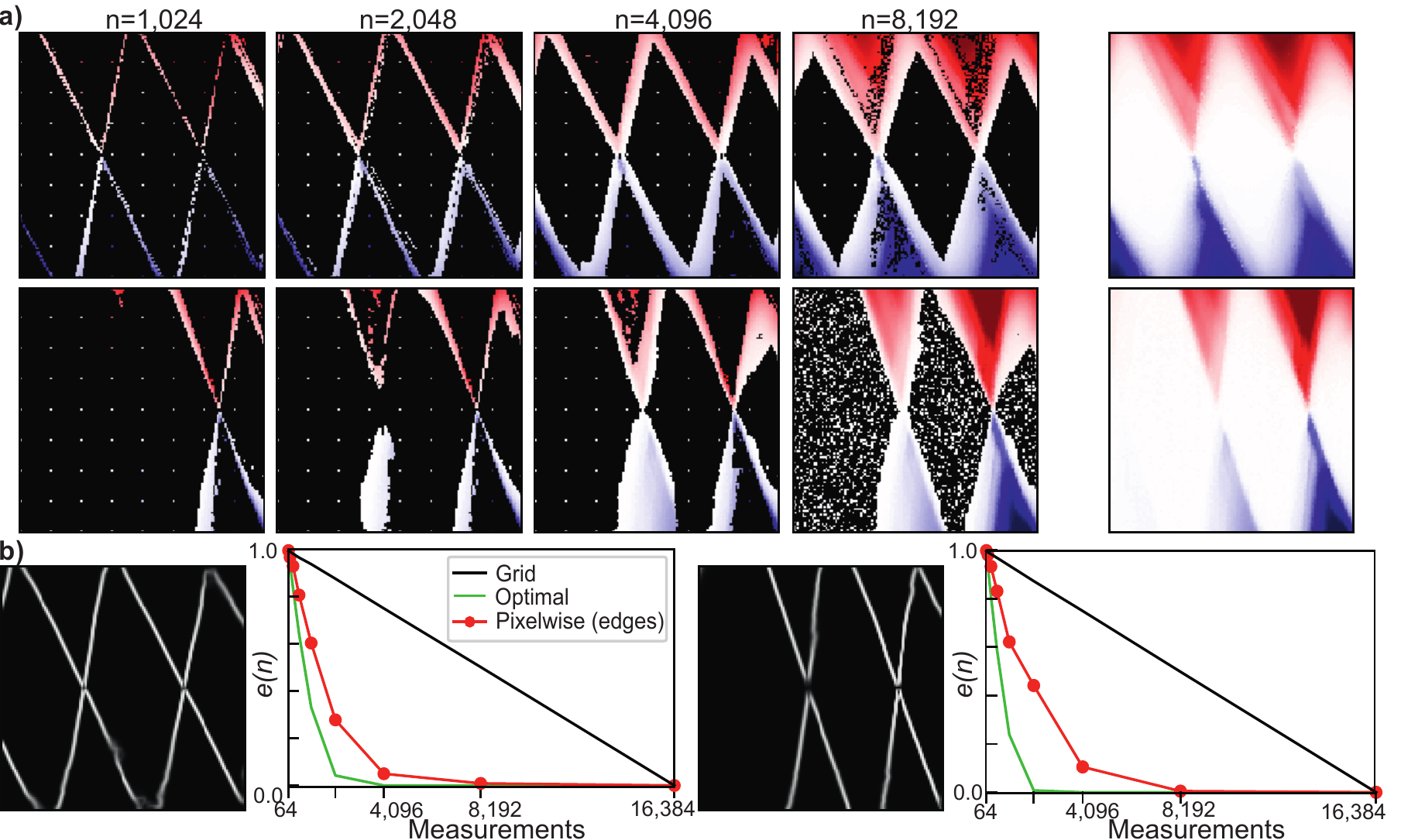}
\caption{\label{fig:supple_acquisition}
Examples of measurements made by the acquisition function that are designed to minimise disagreement between segmentation maps of reconstructions leading to the emergent behavior of measuring edges.
}
\end{figure}
\end{document}